# TEMImageNet Training Library and AtomSegNet Deep-Learning Models for High-Precision Atom Segmentation, Localization, Denoising, and Super-Resolution Processing of Atomic-Resolution Images

Ruoqian Lin[1], Rui Zhang[2], Chunyang Wang[2], Xiao-Qing Yang[1], Huolin L. Xin[2*]

[1] Chemistry Division, Brookhaven National Laboratory, Upton, NY 11973, United States
[2] Department of Physics and Astronomy, University of California, Irvine, CA 92697, United States

*Correspondence should be addressed to H.L.X. (huolin.xin@uci.edu)

## Abstract:

Atom segmentation and localization, noise reduction and deblurring of atomic-resolution scanning transmission electron microscopy (STEM) images with high precision and robustness is a challenging task. Although several conventional algorithms, such has thresholding, edge detection and clustering, can achieve reasonable performance in some predefined sceneries, they tend to fail when interferences from the background are strong and unpredictable. Particularly, for atomic-resolution STEM images, so far there is no well-established algorithm that is robust enough to segment or detect all atomic columns when there is large thickness variation in a recorded image. Herein, we report the development of a training library and a deep learning method that can perform robust and precise atom segmentation, localization, denoising, and super-resolution processing of experimental images. Despite using simulated images as training datasets, the deep-learning model can self-adapt to experimental STEM images and shows outstanding performance in atom detection and localization in challenging contrast conditions and the precision consistently outperforms the state-of-the-art two-dimensional Gaussian fit method. Taking a step further, we have deployed our deep-learning models to a desktop app with a graphical user interface and the app is free and open-source. We have also built a TEM ImageNet project website for easy browsing and downloading of the training data.

## Introduction

In the past decade, the widespread availability of aberration-corrected annular dark-field scanning transmission electron microscopy (ADF-STEM) that offers reliable atomic-scale imaging of materials, has enormously benefited many fields ranging from nanocatalysts and batteries to electronic and structural materials. Using advanced aberration-corrected ADF-STEM, direct acquisition of real-space images with 50-pm resolution can be achieved at high acceleration voltages (300 keV)[1,2]. Recently, Muller and his coauthors have demonstrated that by combining a ptychography technique with a highly sensitive pixelated detector, the resolution envelope can be extended to 39 pm even at low acceleration voltages (80 keV), a condition that can greatly reduce electron beam damage to low-atomic-number materials while retaining ultrahigh resolution[3]. However, acquiring and maintaining these high-resolution instruments incur high costs and to date recording high quality atomic-scale data is still a time-consuming process—high-quality STEM images are not always available, due to many environmental factors, such as scan jittering, temperature fluctuations, stray electromagnetic fields, sample charging and drifting. In non-ideal ADF-STEM images that are contaminated by noise and distortions, the atomic arrangement might still be recognizable by experienced electron microscopists, but some low-contrast atomic details might not be easily detectable by inexperienced operators. Therefore, it is highly desirable to develop a robust method to detect and localize atoms/atomic columns and restore the atomic-scale information in non-ideal ADF-STEM images. Such methods, if available, can greatly reduce misinterpretation, bias, and human errors. It will not only be a valuable tool to student researchers and materials scientists who use ADF-STEM as a tool but also can assist experienced electron microscopists in automated analysis of large datasets.

Atomic column localization and segmentation in atomic-resolution scanning TEM images with high precision and high robustness is non-trivial. Although several algorithms including graph methods[4], clustering methods[5-9], threshold methods[10] and edge detection methods[11] can achieve reasonable performance in pre-defined sceneries, they tend to fall short when noises are strong and interferences are unpredictable. Particularly, for atomic-scale scanning TEM images, to date there is no established algorithm that is sufficiently robust to detect all atomic features when there is large thickness change in an image. For instance, without human supervision, it is non-trivial to localize the dimmer atomic columns on or near the edge/surface of a particle due to the lower contrast and intensities. Herein, we report the development of a training library and a deep learning method that can perform robust and precise atom segmentation, localization, denoising, and deblurring/super-resolution processing of experimental images. Taking a step further, we have deployed our models to a desktop app with a graphical user interface. The app is free and open-source and it is available for download on Github.[12] We have also built a TEM ImageNet project website for searching, browsing, and downloading of the training images and labels[13].

With the availability of affordable high-bandwidth computing hardware, deep learning or deep convolution neural networks (CNNs) that use multilayer artificial neural networks to achieve human-competitive or superhuman accuracy has gained great traction in both the research and commercial application domains[14-16]. Deep learning is now considered the "Holy Grail" for Computer Vision and deep learning models are increasingly being deployed to application areas that utilize object detection, recognition and classification[17,18]. Even though most of the theoretical frameworks for deep learning were developed by the 90s, the deep learning field did not witness a breakthrough or a surge in results until 2011[19,20]. What really has changed the field in the past 5-7 years are the availability of massive labeled data sets, GPU computing, and investments from the IT industry to create open software frameworks for deep learning[21]. The strong suit of deep CNNs is that, given enough training datasets, it can localize and classify features and patterns in images with high accuracy, precision, and robustness. Therefore, it is well poised for the study of ADF-STEM images of interest in this article. For example, Ziatdinov *et al* developed a "weakly supervised" approach and combined it with deep learning to achieve chemical identification and tracking of local transformations in atomic-resolution images of graphene[22]. LeBeau and his coauthors used AlexNet, a version of deep CNNs primarily used for classification, to preprocess $SrTiO_3$ convergent beam electron diffraction patterns and determine crystal thicknesses[23]. Huang and her coauthors used CNNs to locate defects and extract strain fields in *2d* materials[24]. Even though promising, the deployment of deep learning in the STEM imaging field is somewhat slow compared with other fields. The stagnation is partly due to the lack of sufficiently labeled database for training in which all categories of materials, such as crystalline, amorphous and *2d* materials, are considered. Depending on the application, images from all resolvable crystallographic orientations also need to be included in the dataset. The magnitude of the library makes it impossible to collect all images experimentally and label all atomic columns by hand. In addition, training with human labeled data is not always desired because the model's precision and accuracy is ultimately limited by human's error rate.

To solve this problem, we have developed a forward model that incorporates realistic scan and Poisson noises in simulated images. This enables us to synthesize a large number of experimental-like atomic-scale ADF-STEM images from any crystal structures with known atomic coordinates. Using this method, we have developed ADF-STEM image library, also termed as the TEM ImageNet, which include atomic-scale ADF-STEM images of eight materials projected along multiple different orientations, i.e. zone axes. Randomized linear and nonlinear low-frequency background patterns and interferences are added to the ADF-STEM images to improve the robustness of our deep-learning models. To reduce the false-positive rate, we have also included images of clusters and nanoparticles with tapered edge and sharp facets. We have provided a total of 10 types of ground truth labels to train different types of models for tasks like atom segmentation and detection, noise reduction, background removal, and super-resolution processing. Based on our well-labeled TEM ImageNet library, we show that our encoder-decoder-type deep learning models achieve superior performance in atomic column localization, segmentation, noise reduction and deblur/super-resolution processing of experimental ADF-STEM images of crystal structures that were not included in the training library. The precision of our atomic-column localization model can even outperform the state-of-the-art two-dimensional (*2d*) Gaussian fit method. In the meantime, all deep learning models described in this article are released and incorporated in the open-source application, *AtomSegNet*, that is available for download from Github[1]. *AtomSegNet* is intended to become a pre-processing module of a complete TEM image processing workflow that include extensions of materials and zone axis recognition, crystal phase mapping, dislocation and defect detection, atomic counting, strain mapping, etc. The training data sets and labels are available for download, searching and browsing at the project website. [13]

## Methods

In this section, we describe in detail the method used for generating the training data set and then we present the neural network structure and the training strategy.

**Training Datasets**

Recording atomic-resolution ADF-STEM datasets for a large library of crystal structures with known ground-truth labels is an extremely time-consuming project. Even though the high-resolution images with satisfying quality can be obtained regardless of the cost, it is not a trivial task to define the atomic column labels with high precision and accuracy. To mitigate this problem, we create a forward model that can simulate the experimental-like ADF-STEM images of different atomic structures from different crystallographic orientations with realistic noise models. In this way, the ground truth atomic positions are pre-defined. It is also time efficient to create an experimental-like ADF-STEM image set that comprises of a large number of spatial symmetries, atomic arrangements, zone axes, different noise levels and random backgrounds which can greatly improve the robustness of our models.

**Forward model**: in this study, we used a simple linear imaging model which simulates ADF-STEM images by convolving the projected atomic potential of a material with the point spread function (PSF) of a scanning transmission electron microscope. Here, we only use the simplified version of the linear imaging model which disregards the three-dimensional shape of the point spread function because other than reducing contrast, it is a very subtle effect on atomic resolution images.

$$I(x,y) = \iint \sigma(x',y') \, |\Psi(x-x', y-y')|^2 dx'dy'$$
$$= \sigma \otimes \text{PSF}$$

and

$$PSF(x, y; df) = \frac{4\pi^2}{k^2} \left| \int H(\boldsymbol{k}) \exp[-i\chi(\boldsymbol{k}; df) - 2\pi i \boldsymbol{k} \cdot \boldsymbol{r}] d^2 \boldsymbol{k} \right|$$

Here, we opt out using full quantum mechanical methods, such as Bloch-wave and multislice simulations, to simulate images because the simple linear imaging model we employ here is computationally much more affordable. (For STEM simulation, calculating an $N \times N$-pixel image requires $N \times N$'s multislice simulations. However, the computational complexity of our method is equal one single multislice simulation of a very thin sample.) Using the simple linear imaging model, we can render ten thousand 256-by-256 images within minutes whereas even with GPU acceleration it would still take days for the multislice simulation to compute them. For creating a static library, multislice simulation has its merit as it captures most of the scattering physics. However, for on-the-fly local training, the simple linear imaging model is more desirable because of its speed.

In addition, it has been shown that the apparent atomic column positions in ADF-STEM images may not always correspond to the actual atomic positions[25]. This type of quantum phenomena is heavily crystal structure, thickness and orientation dependent. In addition, quantum mechanical simulation offers quantitatively correct column contrast in the simulated images which is one subtlety that can be compensated by other adjustments of the training sets. (The column contrast can be adjusted in our training images by changing the PSF and the background level.) Because our models aim at reporting the apparent positions of the atomic columns, a simple linear imaging model is sufficient.

From the view of generating static libraries for the community, we have created two versions of the same library with one simulated by the simple linear imaging model[26] and one with the multislice simulation[27].

The second part of the forward model is the simulation of realistic noises in the ADF-STEM images. The primary sources of noise of ADF-STEM are the shot/Poisson noise (also known as counting noise) and the scan noise, which we will describe in detail as follows.

**Poisson noise**. For a given pixel, the expected number of incoming electrons is calculated by $n = t_{dwell} \times I/e$. The counted electrons in this pixel follows the Poisson distribution, $P(n) = e^{-n} n^k / n!$ (Because photomultiplier has extremely high quantum efficiency, we ignore the propagation of additional noises.)

**Scan noise**. Random or periodic electromagnetic field or circuit level interreference can cause the beam to deviate away from the expected scanning position; therefore, the effect of the scan noise is a geometrical transformation of the ideal images. We denote the deviation vector by $\boldsymbol{\delta}_{i,j} = (\delta_x^{i,j}, \delta_y^{i,j})$ where $i$ is the row number, and $j$ is the column number. We define the horizontal direction is the fast scanning direction and the vertical direction is the slow scanning direction. Here, for simplicity, we assume that the beam deviation vector does not change when it scans through a horizontal row, i.e. the deviation vector $\boldsymbol{\delta}^{i,j} = \boldsymbol{\delta}^i$ and $\delta_x^i$ and $\delta_y^j$ both follow the same normal distribution modulated by the periodic line frequency, i.e.

$$\delta_{i,x} = f(i|Normal(\mu = 0, \sigma = 1)) \times \sigma_{jitter} \sin(2\pi f t)$$
$$\delta_{i,y} = f(i|Normal(\mu = 0, \sigma = 1)) \times \sigma_{jitter} \sin(2\pi f t + \phi_0)$$

So, the final image is a transformation of the ideal image $I_0$ by

$$I(i,j) = I_0(i - \delta_x^i, j - \delta_y^j)$$

Some example images of how the noise model affect the images are shown in Figure 1.

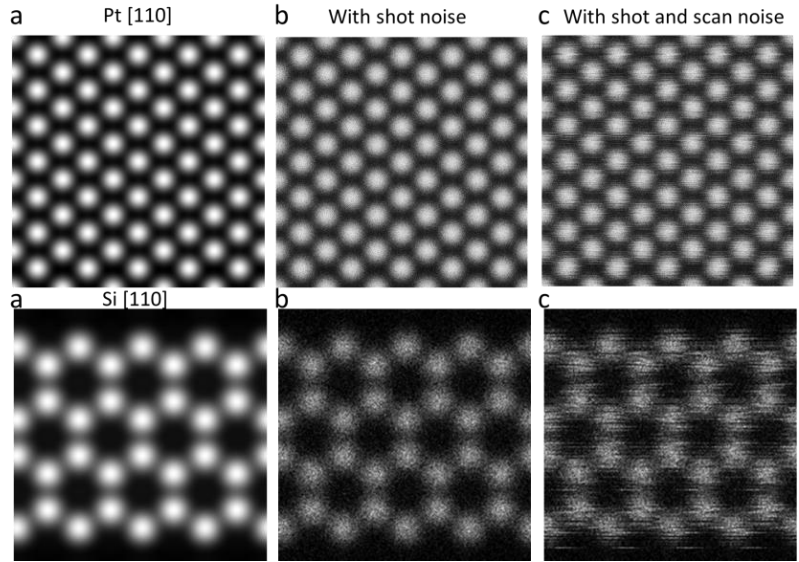

Figure 1. Synthetic images. (a) images created by the simple linear imaging forward model with (b) synthetic shot and (c) scan noises.

**Library composition and augmentation.** To construct a training library with a variety of spatial symmetries, column contrast, and thickness effects, we have included images of the bulk structures of the following materials and orientations: Pt [001], Pt [110], NiO [001], NiO [110], SrTiO$_3$ [001] and [110], DyScO$_3$ [110], Si [110], graphene, amorphous graphene, single-layer MoS$_2$, rutile TiO$_2$ [001], [100] and [110], (Li)CoO$_2$ [010]. We have also included images of the faceted Pt nanocrystal to increase the robustness of finding atoms at boundaries and edges of nanoparticles and interfaces.

To enable robust and scale-free training we have included the following randomized operations in Table 1 in the simulation of the training images and some example images are shown in Figure 2.

Table 1. Augmentation operations

| Operations | Randomization |
|---|---|
| Field of view | 8, 10, 20, 30, 40 angstroms |
| Rotation | 0 to 90 degrees with 15-degree intervals |
| Background | constant, linear ramp, non-linear patterns |
| Shot and scan noise | The shot and scan noise level is randomized |
| Nanoparticle shape | The distance of the 111 facets have been randomized in the dataset |
| Position offset | Randomized |
| Imaging conditions | 1. 200 keV, 24 mrad, source size=0.9 Å, C3/5=0, d$f$=0, $\sigma_{jitter} = 0.2$ Å <br> 2. 100 keV, 30 mrad, source size=0.8 Å, C3/5=0, d$f$=0, $\sigma_{jitter} = 0.2$ Å <br> 3. 200 keV, 10.5 mrad, source size=0.9 Å, C3/5=0, d$f$=0, $\sigma_{jitter} = 0.2$ Å <br> 4. 200 keV, 10.5 mrad, source size=0.9 Å, C3/5=0, d$f$=0, $\sigma_{jitter} = 0.2$ Å <br> 5. 200 keV, 10 mrad, source size=1.6 Å, C3/5=0, d$f$=0, $\sigma_{jitter} = 0.2$ Å |

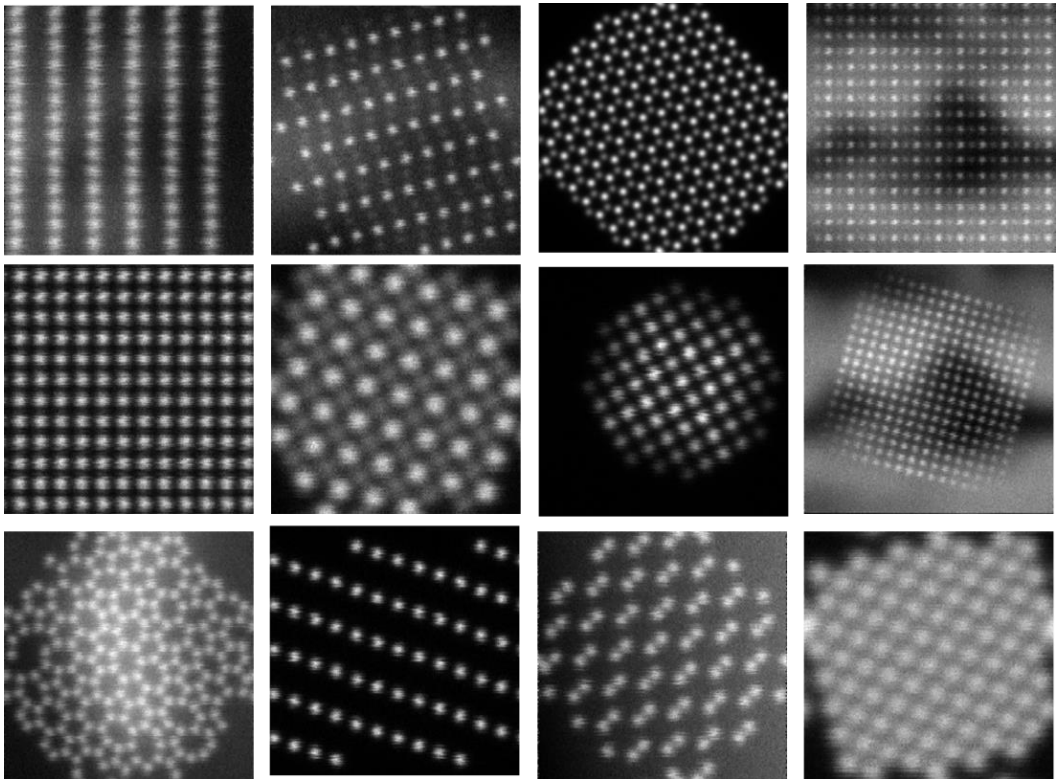

Figure 2. A few examples of the synthetic ADF-STEM images from the training library.

**Ground truth labels.** We have trained our model to perform atom segmentation, atomic-column Gaussian mapping, intensity-preserving super-resolution (deblur) processing, denoising and background removal. Their respective ground truth labels are shown in Figure 3 and Table 2. The width of the circular mask is defined by the full width at half maximum of the point spread function and the width of the Gaussian mask is 0.2 angstrom.

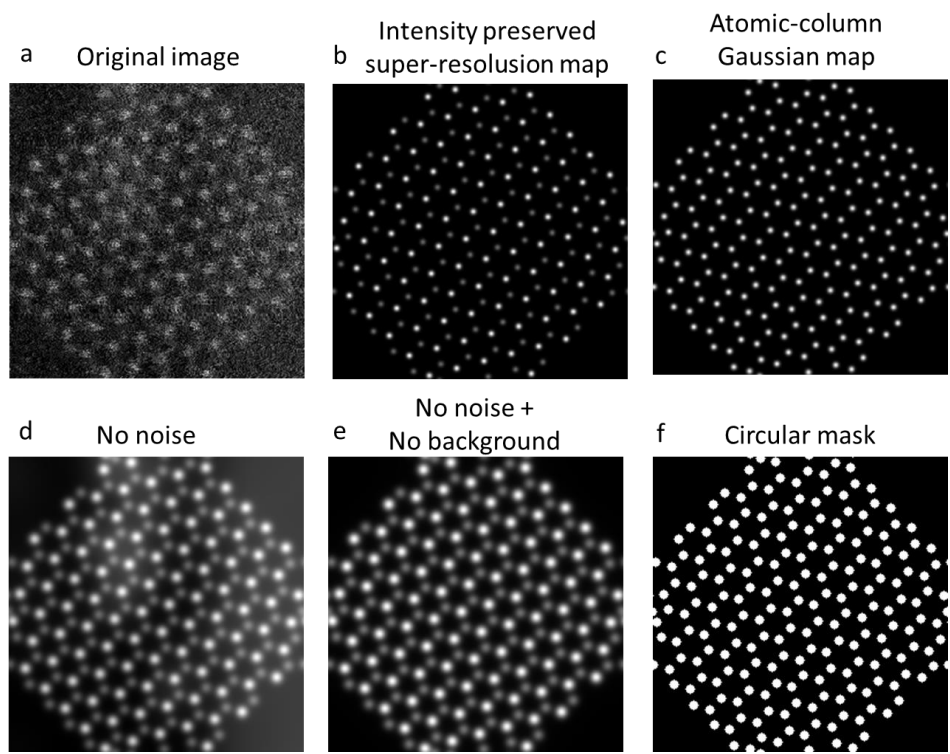

Figure 3. The (a) synthetic image and ground-truth labels for (b) intensity-preserving super-resolution (deblur) processing (c) atomic-column Gaussian mapping, (d) denoising, (e) denoising+background removal, (f) atomic-column segmentation.

Table 2. Description of ground truth labels

| Ground truth label | Application |
| --- | --- |
| **circularMask** | Segmentation labels for atomic column segmentation (radius of the circular mask is defined in radius label) |
| **gaussianMask** | Gaussian-type labels for superresolution localization of atomic columns (the full width at half maximum of the Gaussian function is 0.2 angstrom) |
| **noNoise** | Images without noises for denoising |
| **noBackgrounnoNoise** | Images without noises and backgrounds for denoising and background removal |

| **noNoiseNoBackgroundSuperresolution** | Intensity preserving superresolution images (Gaussian-type) without noises and backgrounds for repersolution procesisng, denoising and background removal |
|---|---|
| **noNoiseNoBckgroundUpinterpolation2x** | Intensity preserving images without noises and backgrounds with 4x more pixels for upinterpolation, denoising and background removal |
| **noNoiseUpinterpolation2x** | Intensity preserving images without noises and 4x more pixels for upinterpolation and denoising |
| **smallcircularMask** | Segmentation labels for atomic column segmentation (radius of the circular mask is defined in smallradius label) |
| **position** | (x, y) positions of each atomic columns in the image |
| **positionRadius** | (x, y, r) positions and radius of each atomic columns in the image |
| **radius** | Radius of the circularMask label is defined as $$r = \sqrt{\left(\frac{0.5\lambda}{\alpha_{max}}\right)^2 + \left(\frac{d_{source}}{2}\right)^2}$$ |
| **smallRadius** | 70% of the radius defined above |

**Network structure**

For the atomic column segmentation, super-resolution/deblur processing, we deployed an encoder-decoder type, U-net architectured CNN network. It has been shown that U-net can work with very few training images and yields precise segmentations for cells tracking tasks[28]. One important feature of U-net is that it concatenates high-resolution feature channels, which directly come from the encoding layers, with the decoding layers to preserve high-resolution context information.

In our model, the contracting path (left side in Figure 4) consists of the repeated application of two 3x3 convolutions, a rectified linear unit (ReLU) and a 2x2 max pooling operation with stride 2 for downsampling; the expansive path (right side in Figure 4) consists of an up-convolution, a concatenation with the corresponding feature map from the contracting path, two 3x3 convolutions and a ReLU.

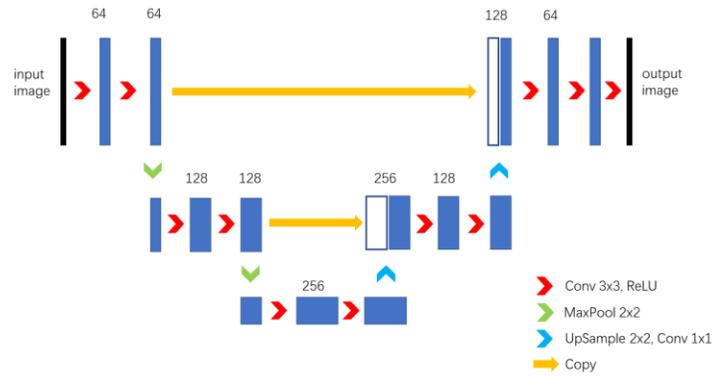

Figure 4. U-net structure in our model.

**Loss function and training strategy**. In our test trainings, we found that mean squared error (MSE) loss function has the tendency to increase false positive rate because the ground truth labels cover a small fraction of the total image area. Therefore, we use a modified chi-square function:

$$\chi^2_{mod} = \sum \frac{\left(I_{i,j} - I_{i,j}^{ground\ truth}\right)^2}{I_{i,j}^{ground\ truth} + \max(I^{ground\ truth})/10}$$

This loss function penalizes false atoms in the background area.

The dataset is split to training set and testing set randomly, with the training set percentage as 75%. The final average training loss after 200 epochs is 0.0174 and the final average testing error is 0.0195. Batch size is 4.

**Atom localization**: Otzu's method is implemented to binarize the atomic features from the map generated by our models[29]. After binarization, each disconnected area is considered an atomic column. The column positions are localized by calculating the geometric centers of the disconnected area. This Otzu's localization method performs the best when couple with outputs from models that were trained on the Circular Mark and Gaussian Mask ground truth labels.

**Benchmark methods**: We use transfer learning to customize a pre-trained faster R-CNN network[30] for direct atomic detection. We have also implemented two-dimensional (*2d*) Gaussian fit to determine atomic column positions. *2d* Gaussian fit is considered the golden method in the transmission electron microscopy (TEM) field for atomic column localization.[31] These two methods are used as baselines to benchmark the precision of the Otzu's atom localizer described above.

## Result & Discussion

**Validation of the AtomSegNet models using TEMImageNet**: To validate our AtomSegNet models, we have performed visual inspection of the performance of various trained models using data from the validation set. Figure 5 shows validation results of the models that were trained to perform super-resolution/deblurring, atom circular segmentation, atomic-column Gaussian mapping, denoising and denoising+background removal using the following labels, `noNoiseNoBackgroundSuperresolution`, `circularMask`, `gaussianMask`, `noNoise`, `noBackgrounnoNoise`.

By applying our models, the centers of the atomic columns can be clearly identified by the following three networks, super-resolution/deblur processing, atom circular segmentation, and atomic-column Gaussian map. The difference between the super-resolution/deblur processing and atomic-column Gaussian mapping is that the deblurred/super-resolved map preserves the column intensity whereas the atomic-column Gaussian mapping equalizes the intensity of all atomic columns. However, all three methods can accurately recognize the atomic columns even in conditions with strong background and noise contamination. Atom localization was performed on the Gaussian maps using the Otsu's method. Its precision benchmarked against baseline methods, Faster R-CNN and *2d* Gaussian fit, will be discussed in the next sub-section.

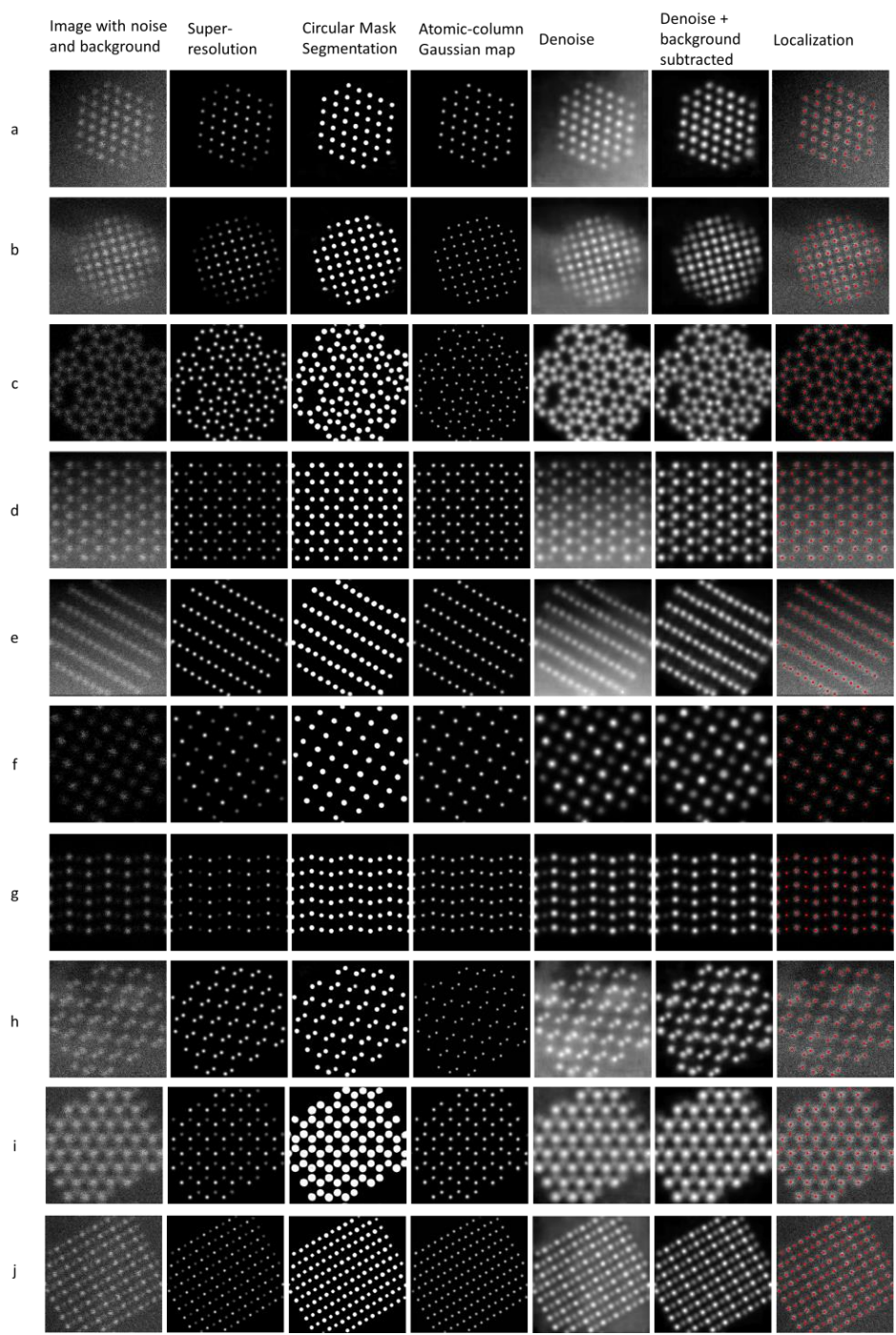

Figure 5. Model validation using the validation set of TEMImageNet. The peak signal-to-noise ratio of images (a)-(j) are around 20 db (or 10 in linear scale).

**Precision of Otsu's atom localizer**. To understand the precision of the different localization method, we compare the Otsu localization method with the *2d* Gaussian fit method and Faster R-CNN. *2d* Gaussian fit is considered the golden method in the TEM field for atomic column localization and faster R-CNN is a deep learning-based method used for direct object detection in 2d images[30]. We have used transfer learning to customize a pre-trained Faster R-CNN to our TEM ImageNet datasets.

    To benchmark the various methods, we have chosen a simulated image with moderately low peak signal-to-noise ratio (PSNR = 20.4 db or 10 in linear scale) and the ground-truth atomic column positions are known. Figure 6a shows the simulated image of a Pt nanoparticle projected along the [110] direction. Figure 6b shows the atomic-column Gaussian map with the Otsu's atom localization results overlaid in the bottom half of the image. Figure 6c shows the Faster R-CNN predicted atomic positions as green boxes. Figure 6d and Figure 6e show the histogram of the deviations of the extracted atomic column positions from of the ground truth labels in *x* and *y* directions respectively. It is easy to see the Fast R-CNN is the least precise (largest $\sigma_x$ and $\sigma_y$). This is because the region proposal network in Faster R-CNN is designed to be fast but precise. Our Otsu's atom localization method coupled with our atomic-column Gaussian map network gives a result that can even outperform *2d* Gaussian fit. The reason that our deep learning method outperforms the traditional *2d* Gaussian fit is likely attributed to the following: *2d* Gaussian fit is based on least square minimization which assumes the noise follows a normal distribution but the dominant noise in the image follows a Poisson distribution. Our neural networks, on the other hand, has learned how to correctly handle the Poisson and the scan noises as well as the priors about the [110]-projected fcc structures. Because the Faster R-CNN method is not sufficiently precise, we only included the Otsu's localization method in an open-source AtomSegNet App[12].

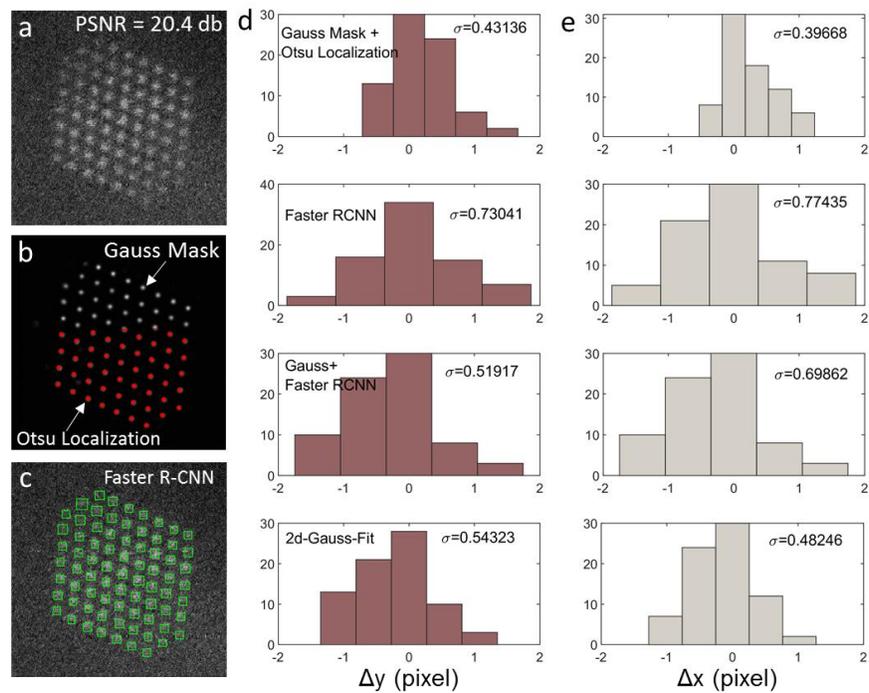

Figure 6. Benchmarking of Otsu's atom localization method against Faster R-CNN and *2d* Gaussian fit.

**Validation of the models on images simulated by the mutlislice method.** We have tested the robustness of our model by applying them to images simulated by the multislice method with aberration and thickness effects are incorporated. As shown in Figure 7, although we have chosen an aberration and defocus condition so that the halo effect is strong, our atomic column Gaussian mapping model faithfully resolve the Sr and Ti atomic columns that are present in the ADF-STEM images.

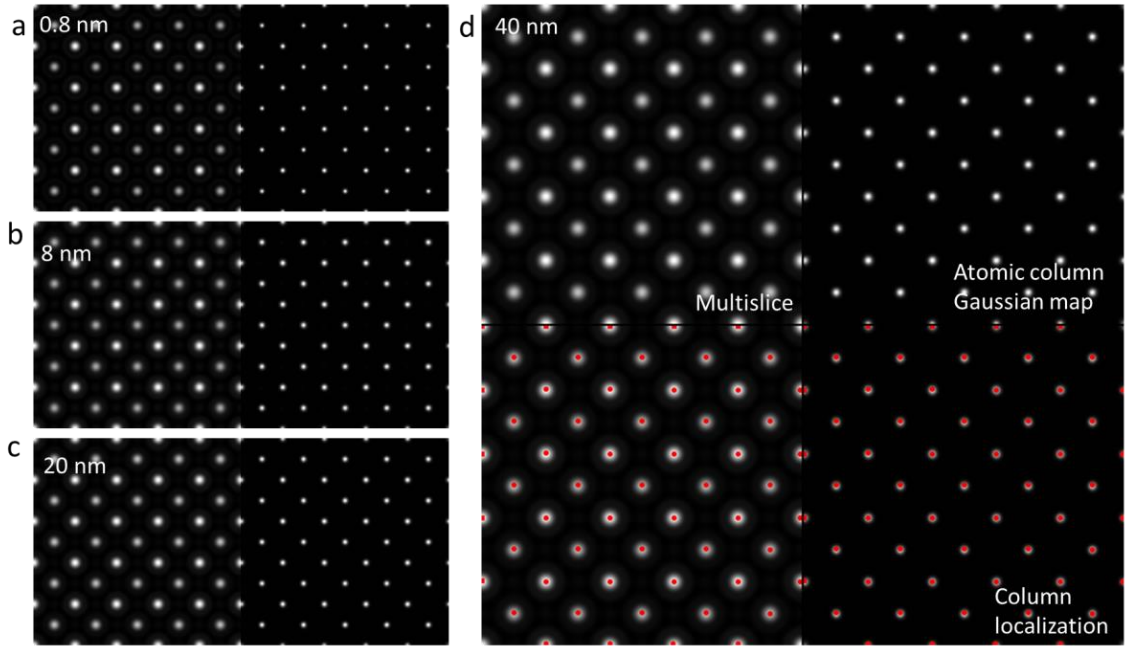

Figure 7. Simulated ADF-SETM images of SrTiO3 along the [001] zone axis and the results processed by the atomic column Gaussian mapping model (200 keV, $α_{max}$ = 20 mrad, df = 80 angstrom, $C_3$ =0.0257 mm, collection angles: 98-196 mrad. Here, the defocus is deliberately chosen slightly off from the optimal value so the halo around the peak is strong).

**Validation of the models on experimental images**. Our model performs well on our validation set, but experimental ADF-STEM images may be interfered by other factors that are not considered in our training sets, for example astigmatism, insufficient resolution, incoherent electron source, and thickness-induced low contrast. Therefore, it is critically important to validate the robustness of our model on experimental images that could contain the these interferences.

*The Known Knowns*: We first apply the AtomSegNet models to ADF-STEM images of periodic crystals, such as DyScO3 [110] (DSO) and silicon (Si) [110]. Because their structures were in the training dataset, they are considered the known knowns. Figure 8a showcases the AtomSegNet processing of a large area of the DSO and Figure 8c shows a magnified area. Figure 8b presents the processed images of Si. For both structures, the atomic columns are detected and localized accurately showing no false positives or false negatives and the denoising results look qualitatively sensible. It demonstrates the robustness of our model when applied to real ADF-STEM images of crystalline materials. It is worth noting that the DSO sample was prepared with Focused Ion Beam and it leaves many redeposited residuals on the surfaces. The background removal model performed well on removing these unwanted low-frequency features. In addition, upon close inspection, Figure 8c shows the image has residual aberrations that renders an asymmetric tail on the Dy columns (the brighter columns). In the `Denoised` image, the tails are still visible as they should be because a denoise process should only remove noise and does not alter the real content of an image. The `donoise+background removal` model, however, corrected the aberrated tails. One can think of it as a process that convolves an unaberrated point spread function with the deblurred/super-resolved map. This is a process that was learned through training.

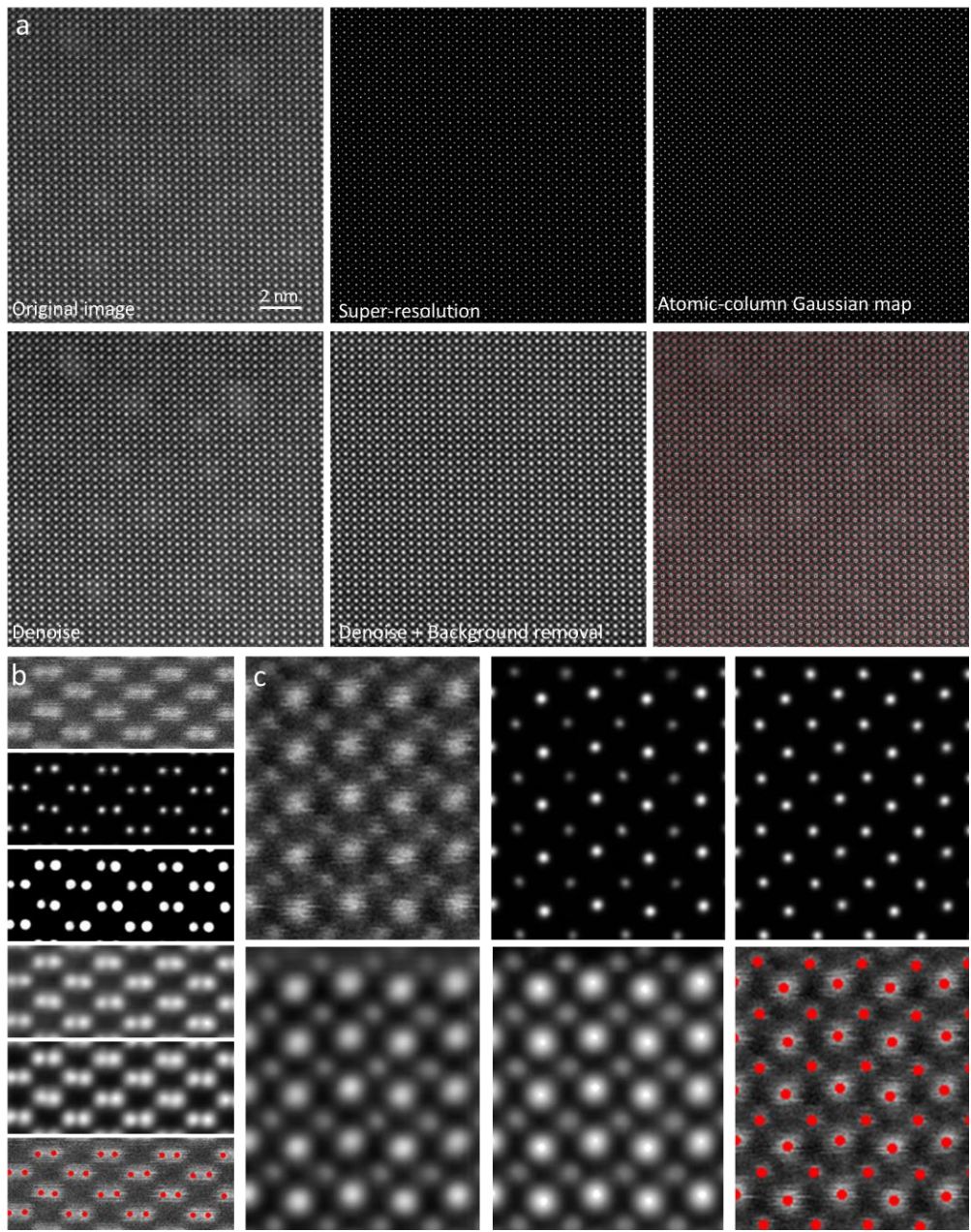

Figure 8. Experimental ADF-STEM images of and the AtomSegNet processing results. (a, c) cross-section specimen of $DyScO_3$ [110], and (b) Si [110].

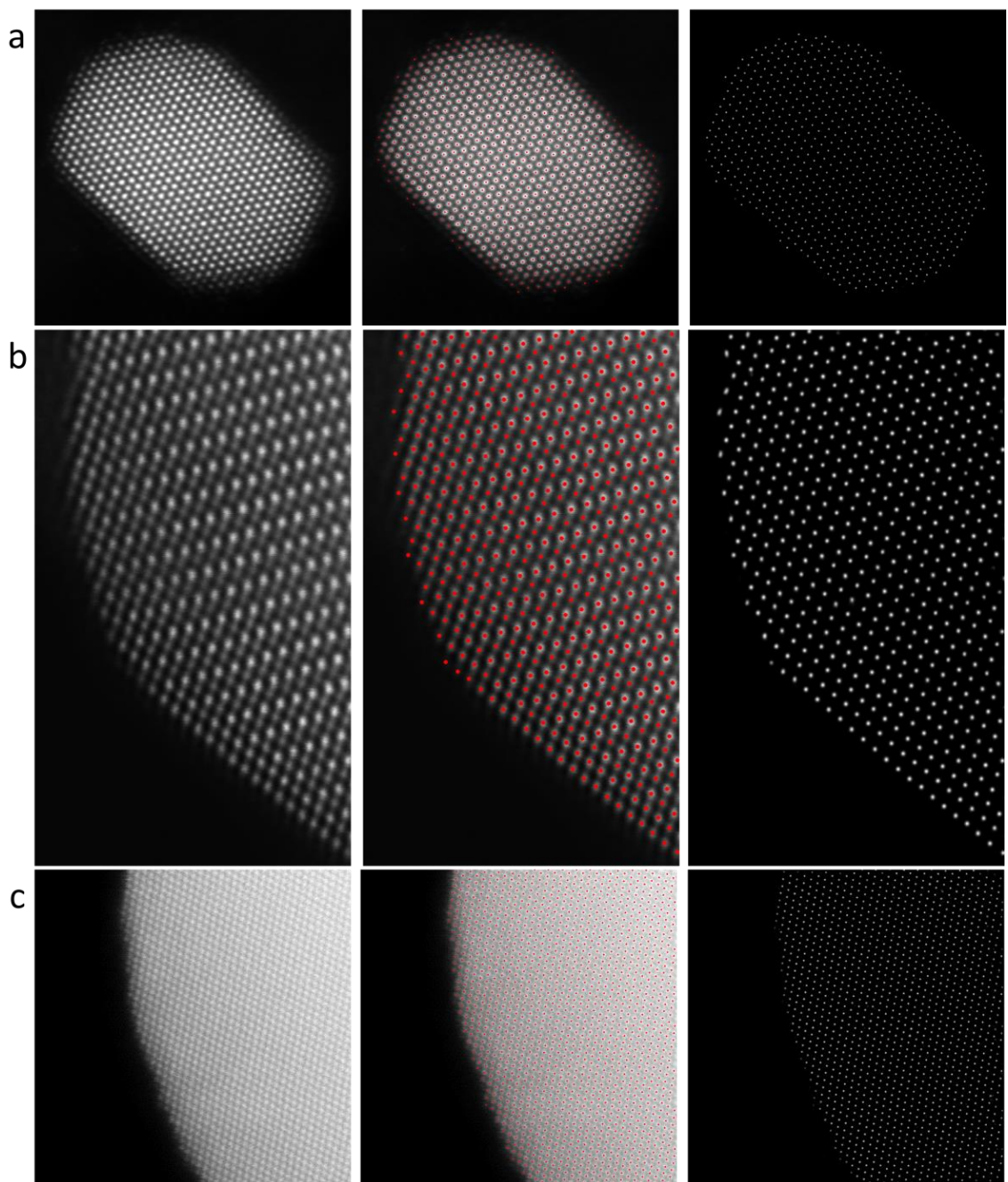

Figure 9. Edge/facet atoms detection/localization. (a) Pt nanoparticle, (b) edge and facet of an PtFe intermetallic nanoparticle, (c) curved edge of an Au particle with large thickness variation.

Another challenging issue in the TEM field is to localize the edge/surface atoms when there is large thickness variation in the recorded images. Observing surface atomic structure is significant for understanding the reaction and degradation mechanism of catalysis and electrode materials since most part of the reaction takes place at the 2~3 atomic layers on the surface. For *2d* materials with uniform thickness, edge atoms detection is straightforward. However in nanoparticle samples, due to the large thickness variation from surface to the bulk interior, the surface atomic columns have lower intensity and are illegible. Herein, we use the noble metal Pt and Au nanoparticles and intermetallic PtFe as examples to demonstrate the capability of our AtomSegNet models for edge/facet atoms detection. It is worth noting that in the ADF-STEM image, the contrast is sensitive to the projected atomic mass of the underlying atomic columns which is commonly referred to as Z-contrast. Therefore, the intensity of PtFe intermetallic atom columns is affected by not only by the thickness variation but also the large atomic mass difference between Pt and Fe. The results shown in Figure 9a and Figure 9b indicate the surface atoms are accurately detected and segmented without ambiguity in both images. Taking a further step, we also tested the capability of edge detection where there is large thickness variation. Figure 9c shows the surface area of an Au nanoparticle with gradually varied thickness, in which the thinner atom columns close to the surface have lower intensities. The result shows that all the surface atom columns are precisely detected and localized without having to choose any hyperparameters. The outstanding results suggest our model is highly robust and capable in edge/facet atom segmentation and localization.

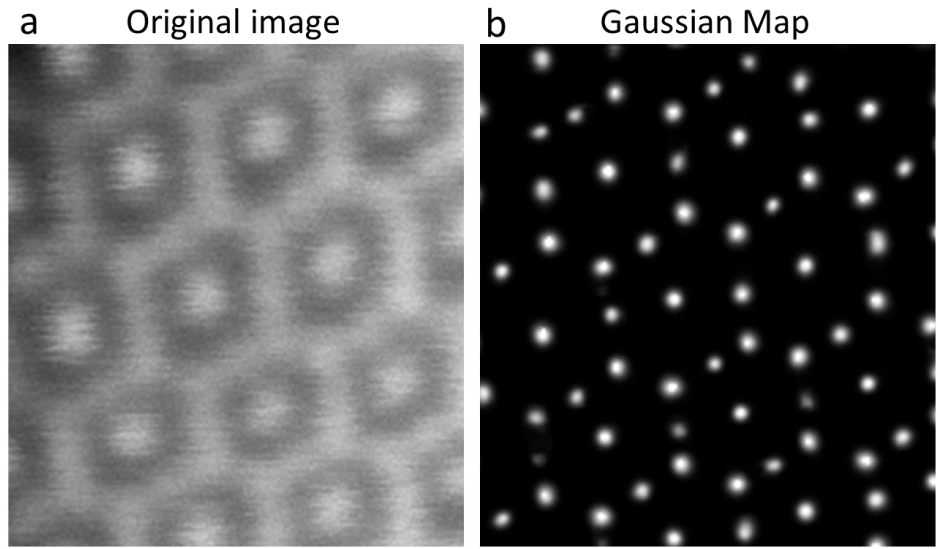

Figure 10. The 'unknown known' test on a spinel structured $Co_3O_4$ material. (a) The ADF-STEM image of $Co_3O_4$. (b) The atomic-column Gaussian map.

**The unknown knowns**: We have tested the performance of the AtomSegNet models on a spinel structured material, $Co_3O_4$. We call this test an 'unknown known' test because this type of pattern was not included in the TEM ImageNet training dataset and the image is under-resolved which is also a condition not included in the training dataset. Figure 10a shows the original under-resolved ADF-STEM image of $Co_3O_4$, in which the adjacent Co atoms are too close to be clearly resolved. Unless the real atomic structure is known in advance, the Co atoms cannot be detected and localized precisely even with the assistance of experienced electron microscopists. Unexpectedly, with our model, the Co atomic columns are accurately recognized in the under-resolved image, corresponding well with the real structure (Figure 10). This result demonstrates our models' 'superhuman' capability to resolve structures and discover the "knowable unknowns".

The reason that our deblurring/super-resolution network exhibit 'superhuman' capability because it essentially learned the regularization from the training images and could perform tasks similar and beyond those formulated in ref[32]. On top of that, it is extremely fast (a few milliseconds of processing time with a GPU) and it does not rely on parameter tuning, i.e. adjusting fitting parameters for the proper point spread function etc.

**Precision analysis on experimental images**: A deep-learning model' accuracy can only be evaluated through comparing the model's outputs with the ground truth labels. For experimental images, however, the ground-truth atomic column positions are not available, not even for crystalline materials of a known type because the atomic column positions in the images are affected by environmental factors such as sample charging, thermal drift and stray fields that cannot be characterized with high precision. However, we can evaluate the precision of our model by measuring the column-to-column spacing, i.e. the *a* and *b* lattice parameters shown in Figure 11e.[31] It is worth noting that the precision measured here is the model's own noise-limited precision compounded by precision loss induced by random distortion. Since the lattice points used for extracting *a* and *b* lattice parameters are spatially close to each other, we assume they share the same random distortions and hence the random errors can be canceled out or minimized.

Figure 11c (i, ii) show histograms and the precision measurement for *x* and *y* lattice parameters extracted by our deep-learning methods and Figure 11c (iii, iv) shows the results extracted by *2d* Gaussian fit on the original ADF-STEM image of Si [110] (Figure 11a). It is shown our deep-learning method report precisions of 7.14 pm in *a* direction and 6.78 pm in *b* direction. These are uniformly better than those the *2d* Gaussian fit method (7.57 pm and 8.25 pm in *a* and *b* respectively). Please note the pixel size of the original ADF-STEM is 9.4 pm and the estimated peak signal-to-noise ratio is 21 db (12 in linear scale). Our Gaussian localization model reaches sub-pixel precision, and the performance is on par to the benchmark measurement in Figure 6.

Another question worth asking is that would our denoise model help improve the precision of atomic column localization. Figure 11b shows the image after processing by our `denoise+background removal` model. We then applied the atomic-column Gaussian map model to the denoised image and the output is shown in the inset of Figure 11b. Figure 11d shows in both *a* and *b* directions, there is a slight improvement in precision for both our deep-learning method and the *2d*-Gaussian fit method. Again, out deep-learning base method reports higher precision than the 2d-Gaussian fit method. In addition, we can use the central limit theorem, i.e. $\sigma_{mean} = \frac{\sigma}{\sqrt{N}} \rightarrow N = \left(\frac{\sigma}{\sigma_{mean}}\right)^2$, to estimate how many images would be needed to achieve 1-pm precision in column localization. Given the $\sigma_a$ is 6.77 pm and the desired $\sigma_{mean}$ is 1 pm, $N$ is equal to 45. It means a series of 45 images is needed to achieve sub-pm precision. This number is on par with the number reported by Voyles and his colleagues[31].

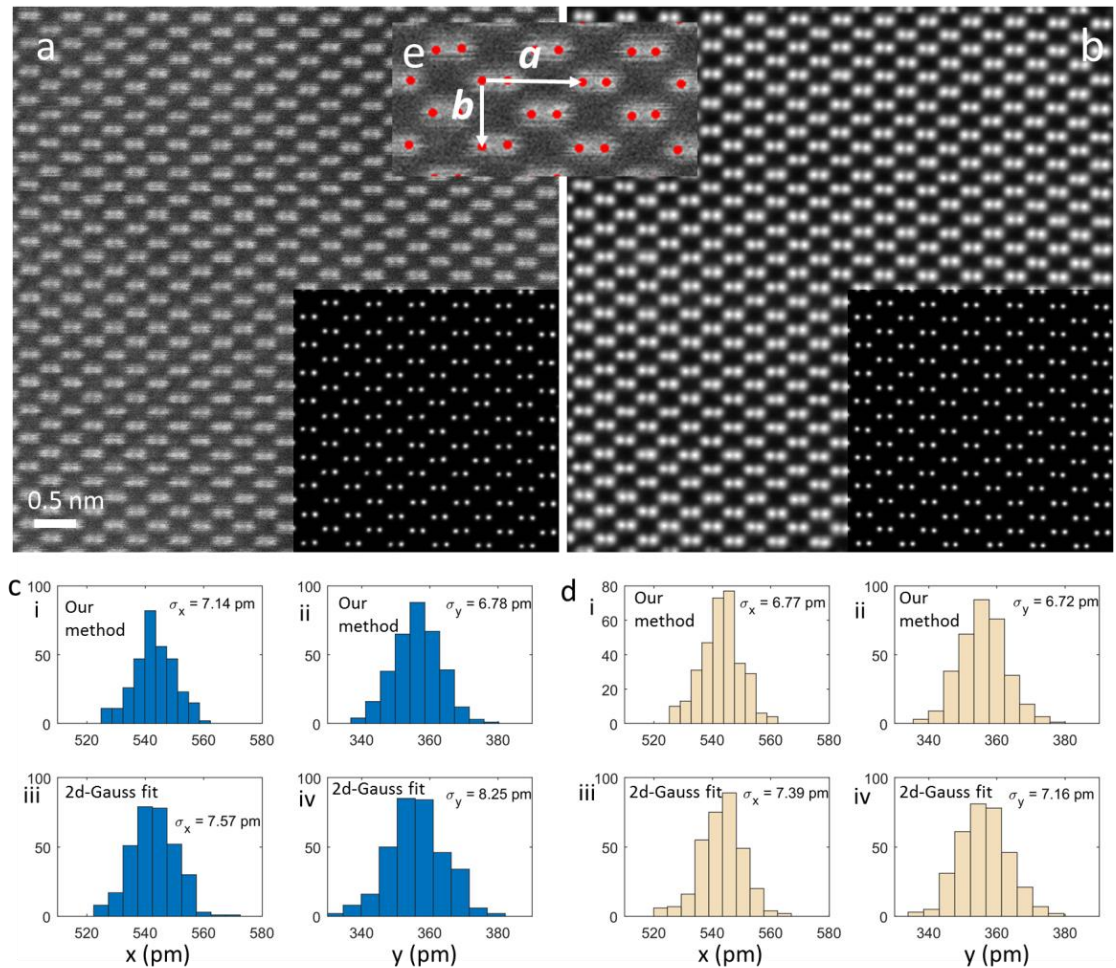

Figure 11. Precision analysis of our deep-learning model on an experimental ADF-STEM image. The pixel size of the image is 9.4 pm and the estimated peak signal-to-noise ratio is 21 db (12 in linear scale). (a) The ADF-STEM image of Si [110]. Inset: the atomic-column

Gaussian map. (b) The image after `denoise+background removal` processing. Inset: the atomic-column Gaussian map. (c and d) The precision analysis. (e) The illustration of the *a* and *b* lattice.

**Discussion**

In this work, we show that using forward modeling based on a linear imaging model, we can rapidly generate near realistic atomic-resolution ADF-STEM images for training many networks with high performance and general usage such as column mapping, deblur/super-resolution processing, denosing, etc. Here, we perform this numerical experiments to verify the effectiveness of the simple imaging model for the following reasons. First, we create this simple forward model envisioning that we will in the future generate training library on-the-fly while the operator is working on the microscope. Our fast and simple forward model can inject priors into a new library much faster than the multislice simulation. Second, the models we desire to train are for column recognition and denoising. Therefore, as long as we can provide a sufficiently diverse library that have different column separation, different coordination structure, and different intensity distribution, the model will be well trained without over fitting the data. Third, although multislice simulation is now much affordable now than before, it is still a computationally intensive task. It is worth noting that, even with the GPU acceleration, it still took us two days to simulate the entire library of images; whereas it would take only 10 mins for a laptop to finish the calculation using our simple linear imaging model. Fourth, our models are scale-free models. It just works like our eyes and brains; it recognizes column patterns regardless of their physical separations. For the ground truth labels, if we use our simple forward model, we can easily rely on the atomic positions as the ground truth label without having to do any refitting to the generated images. Again, from on-the-fly deployment point of view, it is much faster and safer to adopt the single linear imaging method we proposed in this manuscript as it is fool proof and less prone to artifacts. We want to emphasize that our forward model, in terms of column position, is universally correct as it is the thin sample limit of the multiscale method.

**Conclusion**

Detection and localization of atomic columns and the restoration of atomic-scale information in non-ideal ADF-STEM images are highly important for characterizing the atomic structure and understanding the structure-property relationship. However, atom localization through deep learning remains challenging partly due to the lack of sufficiently labeled database for training which is considered an extremely time/cost consuming project. To solve this problem, we create a forward model that can simulate the experimental-like ADF-STEM images. Using this forward model, we have created a TEM ImageNet library composed of training images of different atomic structures from different crystallographic orientations with realistic noise models. By training on this TEM ImageNet library, our deep-learning method can readily self-adapt to the experimental ADF-STEM images and show outstanding robustness in some challenging tasks such as deblurring/super-resolution processing, atom segmentation/localization and edge/surface atom detection. Our models also consistently outperform the precision of the golden method, 2d Gaussian fit, in locating atomic column positions. Furthermore, we have deployed our model to a desktop app with a graphical user interface and the app is free, open-source and available for download on Github.[12] Our model will not only be a valuable tool to researchers and materials scientists but also can assist experienced electron microscopists in automatic analysis of large datasets.

## Acknowledgement

This work is supported by the Early Career Research Award of HLX provided by the Materials Science and Engineering Divisions, Office of Basic Energy Sciences of the U.S. Department of Energy, under award no. DE-SC0021204 (program manager Dr. Jane Zhu). R. Z.'s effort on this project was supported by HLX's startup funding. R. L. and X.-Q. Y. were supported by the Assistant Secretary for Energy Efficiency and Renewable Energy, Vehicle Technology Office of the U.S. Department of Energy through the Advanced Battery Materials Research (BMR) Program, including Battery500 Consortium under Contract DE-SC0012704.

## Author Contributions

HLX conceived the idea. All authors designed the experiment and wrote the manuscript.

## Competing interests

The authors declare no competing interests.

# Data availability

Data and code are available at https://github.com/xinhuolin/AtomSegNet And http://TEMImageNet.com .

# Reference

1  Sawada, H. *et al.* STEM imaging of 47-pm-separated atomic columns by a spherical aberration-corrected electron microscope with a 300-kV cold field emission gun. *Microscopy* **58**, 357-361, doi:10.1093/jmicro/dfp030 (2009).

2  Erni, R., Rossell, M. D., Kisielowski, C. & Dahmen, U. Atomic-resolution imaging with a sub-50-pm electron probe. *Phys Rev Lett* **102**, 096101, doi:10.1103/PhysRevLett.102.096101 (2009).

3  Jiang, Y. *et al.* Electron ptychography of 2D materials to deep sub-ångström resolution. *Nature* **559**, 343-349, doi:10.1038/s41586-018-0298-5 (2018).

4  Peng, B., Zhang, L. & Zhang, D. A survey of graph theoretical approaches to image segmentation. *Pattern Recognition* **46**, 1020-1038, doi:https://doi.org/10.1016/j.patcog.2012.09.015 (2013).

5  Gómez, D. *et al.* Fuzzy image segmentation based upon hierarchical clustering. *Knowledge-Based Systems* **87**, 26-37, doi:https://doi.org/10.1016/j.knosys.2015.07.017 (2015).

6  Tao, D., Li, X., Wu, X. & Maybank, S. J. General Tensor Discriminant Analysis and Gabor Features for Gait Recognition. *IEEE Transactions on Pattern Analysis and Machine Intelligence* **29**, 1700-1715, doi:10.1109/TPAMI.2007.1096 (2007).

7  Wang, L. & Pan, C. Robust level set image segmentation via a local correntropy-based K-means clustering. *Pattern Recognition* **47**, 1917-1925, doi:https://doi.org/10.1016/j.patcog.2013.11.014 (2014).

8  Wang, L., Wu, H. & Pan, C. Region-based image segmentation with local signed difference energy. *Pattern Recognition Letters* **34**, 637-645, doi:https://doi.org/10.1016/j.patrec.2012.12.022 (2013).

9  Liu, T., Tao, D. & Xu, D. Dimensionality-Dependent Generalization Bounds for k-Dimensional Coding Schemes. *Neural Computation* **28**, 2213-2249, doi:10.1162/NECO_a_00872 (2016).

10 Pandey, D., Yin, X., Wang, H. & Zhang, Y. Accurate vessel segmentation using maximum entropy incorporating line detection and phase-preserving denoising. *Computer Vision and Image Understanding* **155**, 162-172, doi:https://doi.org/10.1016/j.cviu.2016.12.005 (2017).

11 Bhandarkar, S. M., Zhang, Y. Q. & Potter, W. D. AN EDGE-DETECTION

TECHNIQUE USING GENETIC ALGORITHM-BASED OPTIMIZATION. *Pattern Recognition* **27**, 1159-1180, doi:10.1016/0031-3203(94)90003-5 (1994).
12      https://github.com/xinhuolin/AtomSegNet.
13      Tang, Z.-K., Xue, Y.-F., Teobaldi, G. & Liu, L.-M. The oxygen vacancy in Li-ion battery cathode materials. *Nanoscale Horizons* **5**, 1453-1466 (2020).
14      Wang, R., Liu, T. & Tao, D. Multiclass Learning With Partially Corrupted Labels. *IEEE Transactions on Neural Networks and Learning Systems* **29**, 2568-2580, doi:10.1109/TNNLS.2017.2699783 (2018).
15      Liu, T., Tao, D., Song, M. & Maybank, S. J. Algorithm-Dependent Generalization Bounds for Multi-Task Learning. *IEEE Transactions on Pattern Analysis and Machine Intelligence* **39**, 227-241, doi:10.1109/TPAMI.2016.2544314 (2017).
16      Liu, T. & Tao, D. Classification with Noisy Labels by Importance Reweighting. *IEEE Transactions on Pattern Analysis and Machine Intelligence* **38**, 447-461, doi:10.1109/TPAMI.2015.2456899 (2016).
17      Sommer, C. & Gerlich, D. W. Machine learning in cell biology – teaching computers to recognize phenotypes. *Journal of Cell Science* **126**, 5529, doi:10.1242/jcs.123604 (2013).
18      de Bruijne, M. Machine learning approaches in medical image analysis: From detection to diagnosis. *Medical Image Analysis* **33**, 94-97, doi:https://doi.org/10.1016/j.media.2016.06.032 (2016).
19      Ciregan, D., Meier, U. & Schmidhuber, J. in *2012 IEEE Conference on Computer Vision and Pattern Recognition.*   3642-3649.
20      LeCun, Y., Bengio, Y. & Hinton, G. Deep learning. *Nature* **521**, 436, doi:10.1038/nature14539 (2015).
21      Oh, K.-S. & Jung, K. GPU implementation of neural networks. *Pattern Recognition* **37**, 1311-1314, doi:https://doi.org/10.1016/j.patcog.2004.01.013 (2004).
22      Ziatdinov, M. *et al.* Deep Learning of Atomically Resolved Scanning Transmission Electron Microscopy Images: Chemical Identification and Tracking Local Transformations. *ACS Nano* **11**, 12742-12752, doi:10.1021/acsnano.7b07504 (2017).
23      Xu, W. & LeBeau, J. M. A deep convolutional neural network to analyze position averaged convergent beam electron diffraction patterns. *Ultramicroscopy* **188**, 59-69, doi:https://doi.org/10.1016/j.ultramic.2018.03.004 (2018).
24      Lee, C.-H. *et al.* Deep Learning Enabled Strain Mapping of Single-Atom Defects in Two-Dimensional Transition Metal Dichalcogenides with Sub-Picometer Precision. *Nano Letters* **20**, 3369-3377, doi:10.1021/acs.nanolett.0c00269 (2020).
25      Hovden, R., Xin, H. L. & Muller, D. A. Channeling of a subangstrom electron beam in a crystal mapped to two-dimensional molecular orbitals. *Physical Review B* **86**, 195415 (2012).

26 *https://github.com/xinhuolin/TEM-ImageNet-v1.3*.

27 *https://github.com/xinhuolin/TEM-ImageNet-Multislice-v1.3*.

28 Ronneberger, O., Fischer, P. & Brox, T. in *Medical Image Computing and Computer-Assisted Intervention – MICCAI 2015.* (eds Nassir Navab, Joachim Hornegger, William M. Wells, & Alejandro F. Frangi) 234-241 (Springer International Publishing).

29 Otsu, N. A Threshold Selection Method from Gray-Level Histograms. *IEEE Transactions on Systems, Man, and Cybernetics* **9**, 62-66, doi:10.1109/TSMC.1979.4310076 (1979).

30 Ren, S., He, K., Girshick, R. & Sun, J. in *Advances in neural information processing systems.* 91-99.

31 Yankovich, A. B. *et al.* Picometre-precision analysis of scanning transmission electron microscopy images of platinum nanocatalysts. *Nature Communications* **5**, 4155, doi:10.1038/ncomms5155 (2014).

32 Fatermans, J. *et al.* Single Atom Detection from Low Contrast-to-Noise Ratio Electron Microscopy Images. *Physical Review Letters* **121**, 056101, doi:10.1103/PhysRevLett.121.056101 (2018).